\documentclass[aps,prapp,reprint,amsmath,amssymb,superscriptaddress,longbibliography]{revtex4-2} 
\usepackage{color,graphicx,float,hyperref,stackengine,braket,bm,mathtools,calc,float,eqnarray}
\usepackage{graphicx}
\usepackage{dcolumn}
\usepackage{colortbl}
\usepackage{enumitem}
\usepackage{placeins}
\usepackage{natbib}

\begin{document}

\title{Wavelet-based Ramsey magnetometry enhancement of a single NV center in diamond}
\author{Ekrem Taha \surname{G\"{u}ldeste}}
\author{Ceyhun \surname{Bulutay}}
\email{bulutay@fen.bilkent.edu.tr}
\affiliation{Department of Physics, Bilkent University, Ankara 06800, Turkey}

\date{\today}

\begin{abstract}
Nitrogen-vacancy (NV) centers in diamond constitute a solid-state nanosensing paradigm. Specifically for high-precision magnetometry, the so-called Ramsey interferometry is the prevalent choice where the sensing signal is extracted from time-resolved spin-state-dependent photoluminescence (PL) data. Its sensitivity is ultimately limited by the photon shot noise, which cannot be sufficiently removed by averaging or frequency filtering. Here, we propose Ramsey DC magnetometry of a single NV center enhanced by a wavelet-denoising scheme specifically tailored to suppress photon shot noise. It simply operates as a classical post-processing applied on a collected PL time series.  Our implementation is based on a method that we named template margin thresholding which enables not only frequency but also time-dependent denoising. We computationally benchmark its DC magnetic field sensing signal-to-noise-ratio improvement over the raw PL data around an order of magnitude.
\end{abstract}

\maketitle

\section{Introduction}
Certain point defects embedded in a solid-state matrix can have optically addressable spin states~\cite{wolfowicz21}. Among these, the negatively charged nitrogen-vacancy (NV$^{-}$) center~\cite{manson18} is particularly favored with its long coherence time up to one second~\cite{abobeih18}, rendering it apt for quantum sensing of fields~\cite{degen08, taylor08, dolde11}, nanoscale magnetic field textures~\cite{casola18}, strain~\cite{barson17}, temperature~\cite{kuscsko13,neumann13}, rotation sensing~\cite{ledbetter12}, and biomedical applications~\cite{li22,aslam23}, as well as three-dimensional mapping of spinful $^{13}$C nuclei in proximity of the defect center~\cite{abobeih19, stolpe23}. 

This initial success has spurred a strong incentive for boosting the sensitivity limits of NV centers~\cite{barry20}. These involve experimental efforts like spin echo time enhancements in shallow NV$^{-}$~\cite{zheng22}, using excited-level anticrossing to enhance readout fluorescence contrast~\cite{steiner10}, or by rising photon collection efficiency~\cite{sage12}, spin to charge conversion~\cite{hopper18,jaskula19}, charge noise suppression for achieving high optical coherence~\cite{kobin23}, nuclear spin polarization for high-field magnetometry~\cite{sahin22}, utilizing double-quantum transitions for doubling the effective gyromagnetic ratio~\cite{mamin14,jarmola21}, to name just a few. Moreover, there are new sensing modalities such as ancilla-assisted repetitive readout~\cite{jiang09,neumann10, lovchinsky16}, or involving proximal nuclear spins~\cite{zhao11,kwiatkowski20,onizhuk23}, covariance magnetometry for obtaining spatiotemporal correlations with two NV$^{-}$ center~\cite{rovny22}, cascading weak measurements into a projective one~\cite{ping23}, in addition to so-called \textit{post-processing} approaches like Bayesian phase estimation protocols proposed for increasing dynamic range and sensitivity~\cite{nusran12,dinani19,zohar23}.

The dominant readout noise in such sensors is primarily mitigated by \textit{averaging} over uncorrelated repeated measurements which attains the so-called standard quantum limit (SQL) as a consequence of the central limit theorem~\cite{braun18, giovannetti04}. This essentially amounts to lowpass filtering of the noise from the sensing signal. A commonly held view to surpass SQL is to harness quantum resources with the most powerful one being entanglement, albeit being very challenging as it makes it prone to faster decoherence~\cite{degen17}. On the other hand, it has been asserted that SQL can be beaten by purely \textit{classical} means as long as \textit{coherence} in the system is preserved during averaging~\cite{braun14,fraisse15}. As a matter of fact, employing noise filtering has proven to be quite potent, as exemplified in boosting gate fidelity to above the surface code error correction limit~\cite{xie23} or coherence protection of NV qubit by deep-learning to improve sensitivity~\cite{xu23}.

Most adversely, NV-center sensors suffer from the photon shot noise (PSN), which has non-stationary temporal characteristics~\cite{degen17,barry20} lowering the effectiveness of conventional frequency filtering. In this respect, more advanced signal processing has been advocated for improving the low signal-to-noise ratio (SNR) of time-resolved spin-dependent fluorescence, to which we will simply refer as photoluminescence (PL)~\cite{gupta16}. Serving for the same purpose an underutilized tool is the wavelet analysis. As a well-established technique~\cite{daubechies92}, it is previously used in this context for the fast detection of temporal magnetic fields~\cite{xu16}, and for resolving spatial positioning of spinful nuclei around central spin~\cite{guldeste22}, and most recently for extraction of both spatially and temporally correlated noise in a two-spin-qubit silicon metal-oxide-semiconductor device~\cite{seedhouse23}. 

In this work, we aim to put forward the efficacy of wavelet-enhanced Ramsey magnetometry of a single NV center, allowing us to develop a time- and frequency-tailored wavelet denoising scheme against PSN. It does not necessitate any quantum resource, and per se, is a post-processing applied over the PL time series that is routinely collected by these sensors. Our implementation, which we term, the template margin thresholding (TMT) method can improve the SNR of PL data significantly in slope detection points which becomes advantageous, especially for limited integration time applications. 

This paper is organized as follows. In Sec.~\ref{hamiltonian}, we describe the spin Hamiltonian and dephasing mechanism of a single NV-center. This is followed by Sec.~\ref{TMT} in which wavelet transform and TMT filter are introduced. Section~\ref{metrics} highlights parameters considered alongside the chosen metrics that underpin the TMT enhancement quantitatively. In Sec.~\ref{results}, we present our numerical simulation results regarding the performance of the TMT filter under different PL settings. Next, in Sec.~\ref{conclusion}, we provide our conclusions and prospects for TMT, and the Appendix contains some basic terminology and information about wavelets, and further technical details on the TMT method.

\section{Model Hamiltonian and dephasing time}\label{hamiltonian}
The ultimate limitation to sensitivity is born out of decoherence, as quantified by the dephasing time $T_2^*$ due to the interaction with the environmental $^{13}$C nuclei forming a spin bath. Choosing the computational $z$-quantization basis parallel to the NV-axis, the general spin Hamiltonian can be split into three parts, 
\begin{equation}
H=H_{\text{NV}} + H_{\text{int}} + H_{\text{env}}, 
\end{equation}
where $H_{\text{NV}}$, represents the NV-center electronic and nitrogen spin, $H_{\text{int}}$ describes the interaction between the NV-center and the environment, and $H_{\text{env}}$ is the Hamiltonian of the spin bath. Under bias magnetic field $\vec{B}_0$, the Hamiltonian terms can be written explicitly~\cite{doherty13},
\begin{align}
  H_{\text{NV}}/\hbar &= DS_z^2 -\gamma_e  \vec{B}_0\cdot\vec{S}  + \epsilon (S_x^2 -S_y^2) \nonumber \\
                      &+ \gamma_N \vec{B}_0\cdot\vec{I} + \vec{I}\cdot\mathbb{P}\cdot\vec{I} + \vec{S}\cdot\mathbb{A}\cdot\vec{I}, \\
  H_{\text{int}}/\hbar &=\sum_{i=1}^N  \vec{S}\cdot\mathbb{A}_i\cdot\vec{I}_i, \label{hint}\\
  H_{\text{env}}/\hbar &= -\gamma_c \vec{B}_0\cdot\sum _{j=1}^{N}\vec{I}_j + \sum_{i<j} \vec{I}_i \cdot \mathbb{D}_{ij}\cdot \vec{I}_j,  \label{henv}
\end{align}
where the sensing-related fields will be separately considered. Here, $D=2\pi \times 2.87$~GHz is zero-field splitting, $S_n$ ($I_n$) represents the electron (nitrogen) spin operator in the $\hat{n}$ direction, $\gamma_e =-2\pi \times 28.024$~GHz/T ($\gamma_N = 2 \pi \times 3.077$~MHz/T) is gyromagnetic ratio of electron (nitrogen), $\epsilon =2\pi \times 100$~kHz is strain induced transverse anisotropy term \cite{jamonneau16}, $\mathbb{P}$ and $\mathbb{A}$ are quadrupolar and hyperfine interaction tensors, respectively. The subscripts in (\ref{hint}) and (\ref{henv}) represent the corresponding spinfull nuclear spin site, so that $\mathbb{A}_i$ refers to the hyperfine interaction tensor for the $i$th nuclear spin. In $H_{\text{env}}$, $\gamma_c =2\pi \times 10.708$~MHz/T is the gyromagnetic ratio of $^{13}$C, and $\mathbb{D}$ is the dipole-dipole interaction tensor. 

To split the degeneracy of $|m_s=\pm1\rangle$ states, we align the bias field, $B_0$, parallel to the NV axis. Choosing this bias field sufficiently large (around a few tens of milliTeslas) also allows us to drop non-secular terms in $\mathbb{A}_i$ interaction tensor ~\cite{maze12, guldeste22}.  We also note that the $^{13}$C spins that are in the vicinity of NV-defect ($<1$~nm) have non-vanishing Fermi-contact terms. It has been shown that they are not the main source of the dephasing mechanism, and since these fastly oscillating terms can be silenced by dynamical decoupling techniques~\cite{barry20}, they can be practically ignored. For a single NV-center, since dephasing time $T_2^*$ highly depends on the hyperfine couplings and hence the spatial distribution of $^{13}$C nuclei~\cite{zhao12, maze12, guldeste18}, we consider randomly distributed 1100 $^{13}$C bath spins, which has approximately 1\% natural abundance in diamond structure. To mimic the thermalized nuclear spin bath dynamics (i.e. with a joint density operator, $\rho \propto \otimes_{i}\hat{\mathcal{I}}_i$), we average out 1000 different initial nuclear spin realizations to calculate coherence of the central spin, under bias field $B_0=40$~mT. With these parameters and via second-order generalized cluster correlation expansion (gCCE)~\cite{yang20, onizhuk21}, we have extracted according to the expression $e^{-(t/T_2^*)^p}$, the dephasing time of a single NV-center, $T_2^*=3.9~\mu$s, and decay factor, $p=2$.

With the help of the bias magnetic field, $B_0$, which sets $|m_s=-1\rangle$ state to be off-resonant, we obtain an isolated two-level system with splitting $\omega_0=D-\gamma_e B_0$ between the states $|m_s=0\rangle$ and $|m_s=1\rangle$. Next, including the sensing and calibration fields, the Hamiltonian in the rotating frame at $\omega_0$ acquires the simple form $H_{\text{sensor}}/\hbar = -\gamma_e [B_{\text{calib}} + B_{\text{sense}}]\sigma_z /2$ where  $\sigma_z$ is the Pauli spin-$z$ operator, $B_{\text{calib}}$ and $B_{\text{sense}}$ are calibration and sensing fields, respectively.  The microwave fields used for initialization and readout in the standard Ramsey magnetometry pulse scheme are omitted ($\pi/2-\tau-\pi/2$)~\cite{degen17}. 

We should make a clear distinction among $B_0$, $B_{\text{calib}}$ and $B_{\text{sense}}$ for the purposes of our simulations. Firstly, we utilize $B_0$ as a bias field to split the degeneracy between $|ms=\pm1\rangle$.  The second field, $B_{\text{calib}}$ which we choose to be $100~\mu$T, has a dual purpose: (\emph{i}) it is used to calibrate the TMT-filter order in the absence of $B_{\text{sense}}$, (\emph{ii}) it controls the number of fringes and hence the slope detection points (SDs) appearing in the preset duration of the PL stream.  As a side note, in practice,  applying $B_{\text{calib}}$ would be equivalent to sampling PL data with a digital down converter. Such an apparatus when the local oscillator is detuned by $|\gamma_e|B_{\text{calib}}$ from the natural frequency of the NV-center, $\omega_0$, provides the same PL profile for SD. Lastly, $B_{\text{sense}}$ is the field that we need to estimate from the shift in the PL stream at the SD points. In the performance assessment part of our work, we set $B_{\text{sense}}$ to a known value of $0.5~\mu$T.  As such, we operate in the regime where $B_{\text{sense}} \ll B_{\text{calib}} \ll B_0$.

\begin{figure*}[!ht]
  \includegraphics[width=1.5\columnwidth]{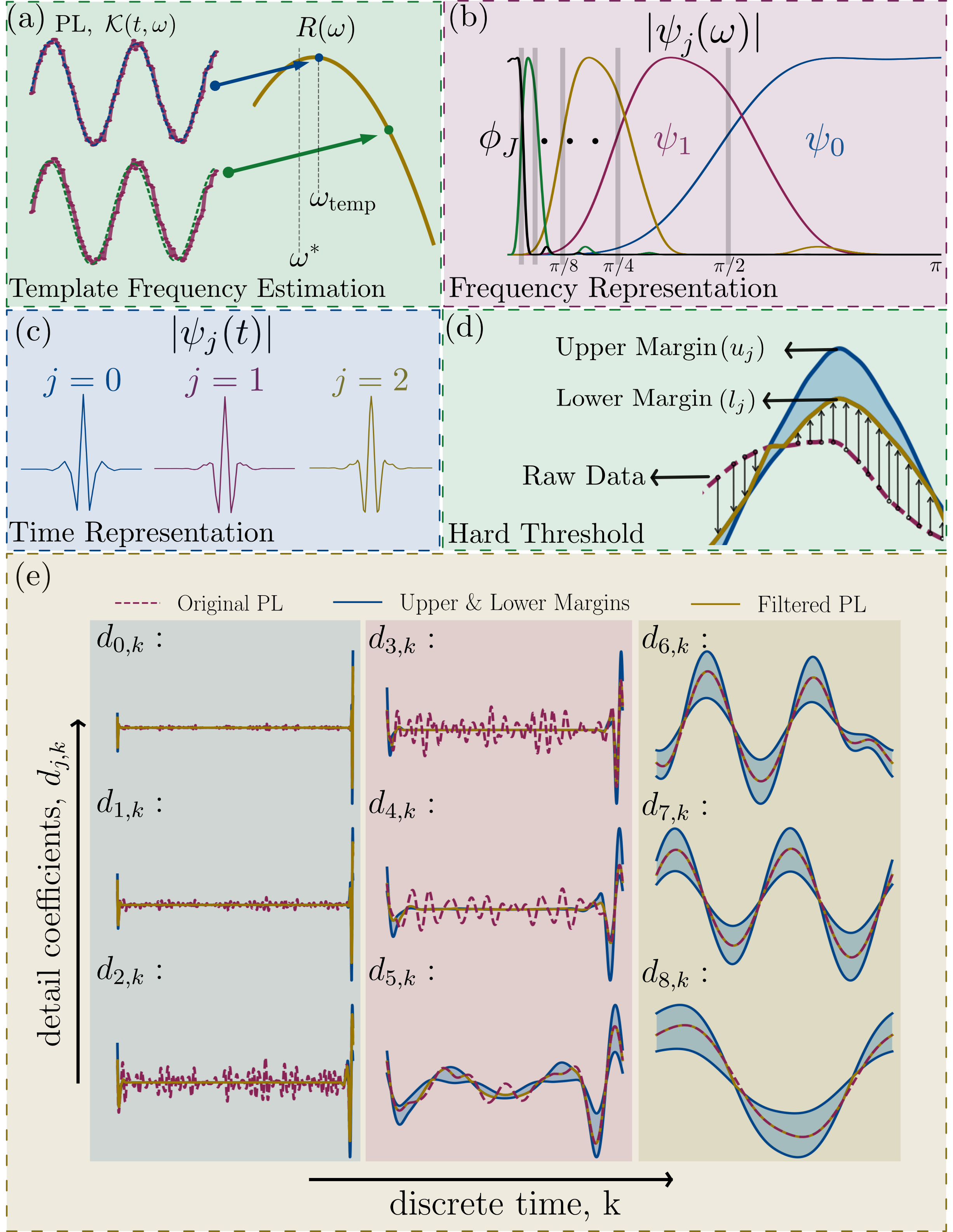}
  \caption{(a) Estimation of template frequency by means of
  Eq.~$(\ref{xcorr})$. The red line represents PL collected between the time
interval $[t_i=0.97~\mu$s,  $t_f=1.75~\mu$s] with sampling frequency
$f_\text{sample}\sim 128$~MHz and $M=25$K, blue and green lines are
$\mathcal{K}(t, \omega)$ at different trial frequencies. The green line
mismatches with the PL, whereas the blue line maximizes the $R(\omega)$ and is
employed as template frequency. There is an offset between the true value
($\omega^*$) and the template ($\omega_{\text{temp}}$). (b) Frequency
representation of \textit{biorthogonal 6.8} wavelet at different scales,
$|\psi_j(\omega)|$, and father wavelet (scaling function) $|\phi_J|$. Each
wavelet $\psi_j$ splits the remaining frequency axis into two and $\phi_J$
spans the lowest frequencies around the $\omega=0$ value to provide a compact
representation of a given signal. (c) \textit{biorthogonal 6.8} wavelet at
different scales in the time domain. (d) Representation of hard threshold (Eq.
$(\ref{hard})$), blue line corresponds to upper and lower margins, red dashed
and yellow lines indicate raw and filtered signal, respectively. (e)
Enhancement of PL signal provided in (a) in the wavelet domain by hard
thresholding. The horizontal axes represent discrete (sampled) time, for all
detail levels from $j=0$ up to $j=8$, and the vertical axes designate detail
coefficients at each level.}\label{algo}
\end{figure*} 

\section{Photon shot noise tailored wavelet enhancement}\label{TMT}
Before we introduce our wavelet-based denoising scheme, we would like to give brief information on the wavelet transform, leaving further details in the Appendix sections.
\subsection{Wavelet transform}
A given signal, $x(t)$, can be represented in the wavelet basis up to the desired decomposition level $J$ as,
\begin{equation}
  x(t) = \sum_{k=-\infty}^{\infty}c_{J, k}\phi_{J,k}(t) + \sum_{j=0}^{J}\sum_{k=-\infty}^{\infty} d_{j,k}\psi_{j,k}(t),
\end{equation}
where the coefficients $c_{J, k}$ and $d_{j,k}$ are called approximation and detail coefficients which can be calculated from the inner product of the signal $x(t)$ with scaling $(\phi_{J})$ and wavelet functions $(\psi_{j,k})$, respectively.  Instead of evaluating the \textit{approximation} and \textit{detail coefficients} with such inner products each time, the discrete wavelet transform (DWT) of $x(t)$ can be calculated by utilizing lowpass and highpass filters recursively (see Appendix C for details) \cite{mallat89, walker08}.  DWT is an efficient method for signal decomposition yet, it suffers from time invariance due to downsampling. In this work, we use an undecimated wavelet transform (UWT), also known as the stationary wavelet transform, which is a DWT without decimation.

\subsection{Template margin thresholding}
We simulate a PL signal collected from an NV sensor, which obeys the Poisson statistics for the spin-dependent readout such that the average number of photons collected per experiment is assumed to be $n_0$ and $n_1$ depending on the projection to the designated qubit states $|0\rangle$ and $|1\rangle$, respectively \cite{barry20}. PL is acquired for some finite time duration starting from $t_i$ to $t_f$ in the presence of a calibration field, $B_{\text{calib}}$ along with a sensing field $B_{\text{sense}}$ we would like to estimate. 

To construct the margins of the wavelet filter, we first extract a so-called template frequency from the PL stream, $N^{\text{raw}}(t)$ through the overlap
\begin{equation} \label{xcorr}
  R(\omega) = \int_{t_i}^{t_f}  N^{\text{raw}}(t)~\mathcal{K}(t, \omega) dt,
\end{equation}
where $\mathcal{K}(t, \omega)$ has the form,
\begin{equation} \label{template}
  \mathcal{K} (t, \omega) =  \dfrac{1}{2}\left[ 1 + \cos(\omega t) e^{-(t/T_2^*)^p} \right](n_0-n_1) + n_1.
\end{equation} 
which represents the \textit{template} PL waveform~\cite{nusran12}. Thus, by searching over $\omega$, it is possible to find a so-called \textit{template frequency}, $\omega_{\text{temp}}$, which maximizes $R(\omega)$, i.e., $d R(\omega) / d\omega|_{\omega=\omega_{\text{temp}}} =0$ as indicated in Fig.~(\ref{algo})~(a). In $(\ref{xcorr})$ we remove the DC offset of both $N^{\text{raw}}(t)$ and $\mathcal{K}(t, \omega)$ to identify the overlap between them, and the exponential part in $(\ref{template})$ represents the qubit dephasing.

The next step is to identify time-dependent margins of the TMT filter according to the photon fluctuations characterized by the typical analytical form \cite{barry20},
\begin{widetext}
\begin{equation}\label{DN}
  \Delta N (t) = \sqrt{\dfrac{(n_0-n_1)}{4} \sin^2(\omega_{\text{temp}}t) + n_0\cos^2{(\omega_{\text{temp}}t/2)} + n_1\sin^2{(\omega_{\text{temp}}t/2)}}.
\end{equation}
\end{widetext}
Hence, we introduce the upper and lower margins which are going to be mapped to the wavelet domain (see Appendix D),
\begin{align}
  \mathcal{U}(t) &= \mathcal{K}(t, \omega_{\text{temp}}) + 10^{-\beta}\dfrac{\Delta N(t)}{\sqrt{T_I M  f_{\text{sample}}}},\label{U}\\
  \mathcal{L}(t) &= \mathcal{K}(t, \omega_{\text{temp}}) - 10^{-\beta}\dfrac{\Delta N(t)}{\sqrt{T_I M  f_{\text{sample}}}},\label{L}
\end{align}
where, we term $\beta$ as the \textit{filter order}, $T_I=t_f-t_i$ is the PL interval, $M$ is repetition number, and $f_{\text{sample}}$ is the PL sampling frequency. These time-budget parameters are explicitly retained in these equations as the shot noise diminishes with the square root of the total integration time \cite{degen17, barry20}.  The upper and lower margins defined in equations $(\ref{U})$ and $(\ref{L})$ can be mapped into the wavelet domain up to a certain detail level along with the raw PL data for shrinking as indicated in Fig.~$\ref{algo}$ (d) (see Appendix D for more details).

The advantage of wavelet transform over conventional frequency filtering is that the former allows time-dependent margins for noise reduction which can be observed with shaded blue regions in Fig.~$\ref{algo}$~(d) and (e). Detail coefficients at a particular level contain only a specific range of frequencies as displayed in Fig.~$\ref{algo}$~(b), so that, the decomposition of PL  with a wavelet at a selected scale provides a simultaneous time-frequency representation. 

In general, the signal values above a certain threshold level are assumed to be the noiseless part of the signal, whereas, the segments that lie below the threshold level are regarded as the noise in the wavelet domain. To get rid of this noise part, depending on the shrinkage type chosen, the segments that are identified to be noise are reassigned, say to zero or a certain value.  In TMT-shrinkage, at each detail level available, we reassign the overflowing raw PL values to the nearest margin value and construct a filtered PL as shown by the yellow lines in Fig.~$\ref{algo}$~(d) and (e). Finally, we map the thresholded signal back to the time domain by inverse UWT and obtain a TMT-enhanced PL, which we represent by $N^{\beta}(t)$ where superscript $\beta$ denotes the order of TMT filter. We would like to note that the time dependence of enhanced (raw) PL, $N^{\beta}$  ($N^{\text{raw}}$)  is sometimes omitted for the sake of simplicity in the remaining part of this paper.

A crucial technical detail is that the wavelet type needs to be chosen according to the amount of \textit{support} at different decomposition levels.  As an example, we show the upper and lower margins in Fig~\ref{algo} (e), for detail levels  $d_{0, k}$ to $d_{4, k}$, where PL is fully dominated by shot noise since $\omega_{\text{calib}}$ is highly off-resonant with the associated wavelet frequencies.  Therefore, choosing a wavelet that decomposes the template PL, $\mathcal{K}$, with mostly vanishing components in these detail levels, is desirable for efficient filtering because in that case, upper and lower margins diminish to very small values.  On the other hand, \textit{on-resonant} components such as $d_{6, k}$ and $d_{7, k}$ in Fig.~\ref{algo} (e), should have relatively large margins for preventing filter to operate inadvertently. To meet these objectives, we opted for the so-called \textit{biorthogonal 6.8} wavelet~\cite{mallat09} to implement the TMT method.

\begin{figure*}[tb]
  \begin{center}
  \includegraphics[width=2.0\columnwidth]{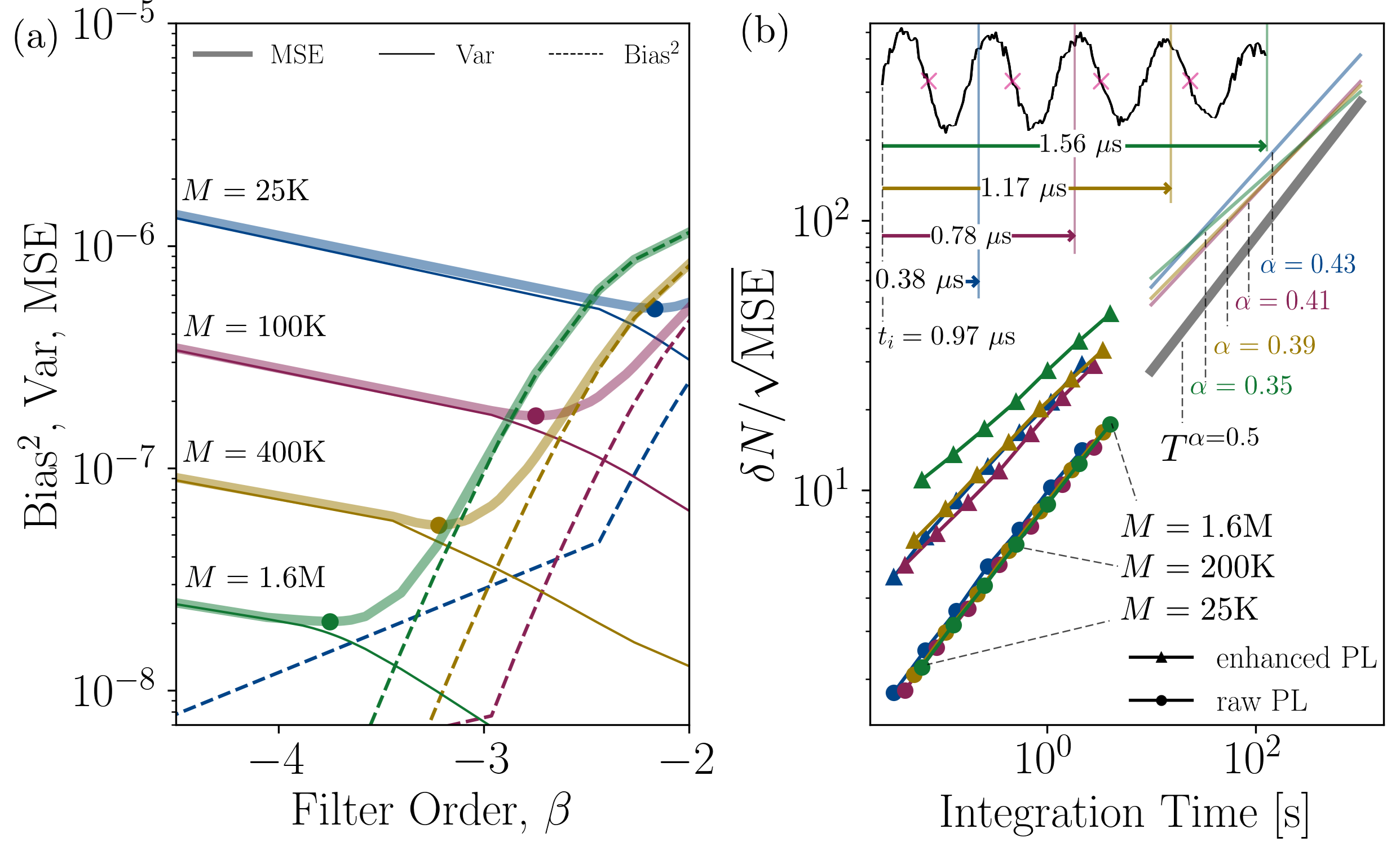}
  \caption{(a) Bias-variance trade-off with respect to filter order for PL taken between $[t_i=0.97~\mu$s,  $t_f=2.14~\mu$s] at  $f_{\text{sample}} \sim 128$~MHz including 3 $n_{\text{SD}}$ points. The MSE minima (i.e., optimum filter orders) are identified, with dots. (b) SNR with respect to total integration time with different PL durations. Dots represent the original PL SNR, and triangle markers represent TMT-enhanced PL data. Each marker at the same line stands for a specific $M$ number ranging from 25K to 1.6M with doubling $M$ numbers each time. Colors represent different PL durations, each starting from  $t_i=0.97~\mu$s, to various $t_f$ values (blue, red, yellow, and green lines have $t_f$ values 1.36~$\mu$s, 1.75~$\mu$s, 2.14~$\mu$s, 2.53~$\mu$s, respectively). All PL time series are collected at $f_{\text{sample}}\sim 128~$MHz.  All data points in (b) are generated at the corresponding optimum filter order.}\label{SNR}
  \end{center}
\end{figure*}

\section{Performance metrics}\label{metrics}
We use PL signals with readout contrast, $C=0.2143$, and with the average number of photons observed per repetition $n_{\text{ave}}=0.196$ based on experiments~\cite{shields15}.  Since the number of photons collected per measurement is very low and hence largely uncorrelated, we assume that the photon statistics can be obtained from the Possionian distribution~\cite{barry20}. We designate with $M$ the number of repeated measurements to obtain each \textit{single} data point over the PL signal. The PL signal is collected over some finite time duration $T_I = t_f-t_i$ so that the total number of data points is given by $T_I f_{\text{sample}}$.  Furthermore, we use $M t_f$ as the integration time, which corresponds to the total time to collect the last data point of a PL signal.

A common metric to quantify the performance of an estimator, such as the TMT enhancement of PL is the mean squared error (MSE). Its statistical assessment necessitates an extra round of $n_{\text{exp}}$ independent experiments as, 
\begin{equation}
\text{MSE} [ N^{\beta}(t_q) ] = \dfrac{1}{n_{\text{exp}}} \sum_{i=1}^{n_{\text{exp}}}  [N_{i}^{\beta}(t_q) - N^{*}(t_q)]^2, \label{mse}
\end{equation}
where $\text{MSE} [ N^{\beta}(t_q) ]$ represents the MSE of TMT enhanced PL, $N^{*}(t_q)$ is the true value for the number of photons, and the $t_q$ refers to the time at $q^{\textit{th}}$ detection point. There are $q=1,\dots, n_{\text{SD}}$ such detection points all selected close to the negative slopes of the PL fringes, as marked by red crosses in the inset of Fig~\ref{SNR} (b).  The MSE in Eq.~($\ref{mse}$) can be decomposed in terms of bias and variance as,   
\begin{equation}
 \text{MSE} [N^{\beta} (t_q)] = \text{Bias}[N^{\beta}(t_q)]^2 + \text{Var}[N^{\beta}(t_q)],
\end{equation}
where,
\begin{align}
  \text{Bias}[N^{\beta}(t_q) ] &= \sum_{i} [ N_{i}^{\beta}(t_q) - N^{*}(t_q)]/n_{\text{exp}}, \\
  \text{Var}[N^{\beta}(t_q) ] &= \sum_{i} [ N_{i}^{\beta}(t_q) - \overline{ N_{i}^{\beta}}(t_q) ]^2 /n_{\text{exp}},
\end{align}
and the overline represents the sample mean. For our results, we use $n_{\text{exp}}=2000$, and consider PLs with various $n_{\text{SD}}$ numbers ranging from one to nine. To facilitate comparing PL streams of varying durations, we introduce fringe-averaged MSE values as, $\text{MSE} [N^{\beta}] = \sum_{q=1}^{n_\text{SD}}\text{MSE} [N^{\beta}(t_q)] / n_{\text{SD}}$.

\section{Results and discussion}\label{results} 

\begin{figure}[tb]
  \begin{center}
  \includegraphics[width=1.0\columnwidth]{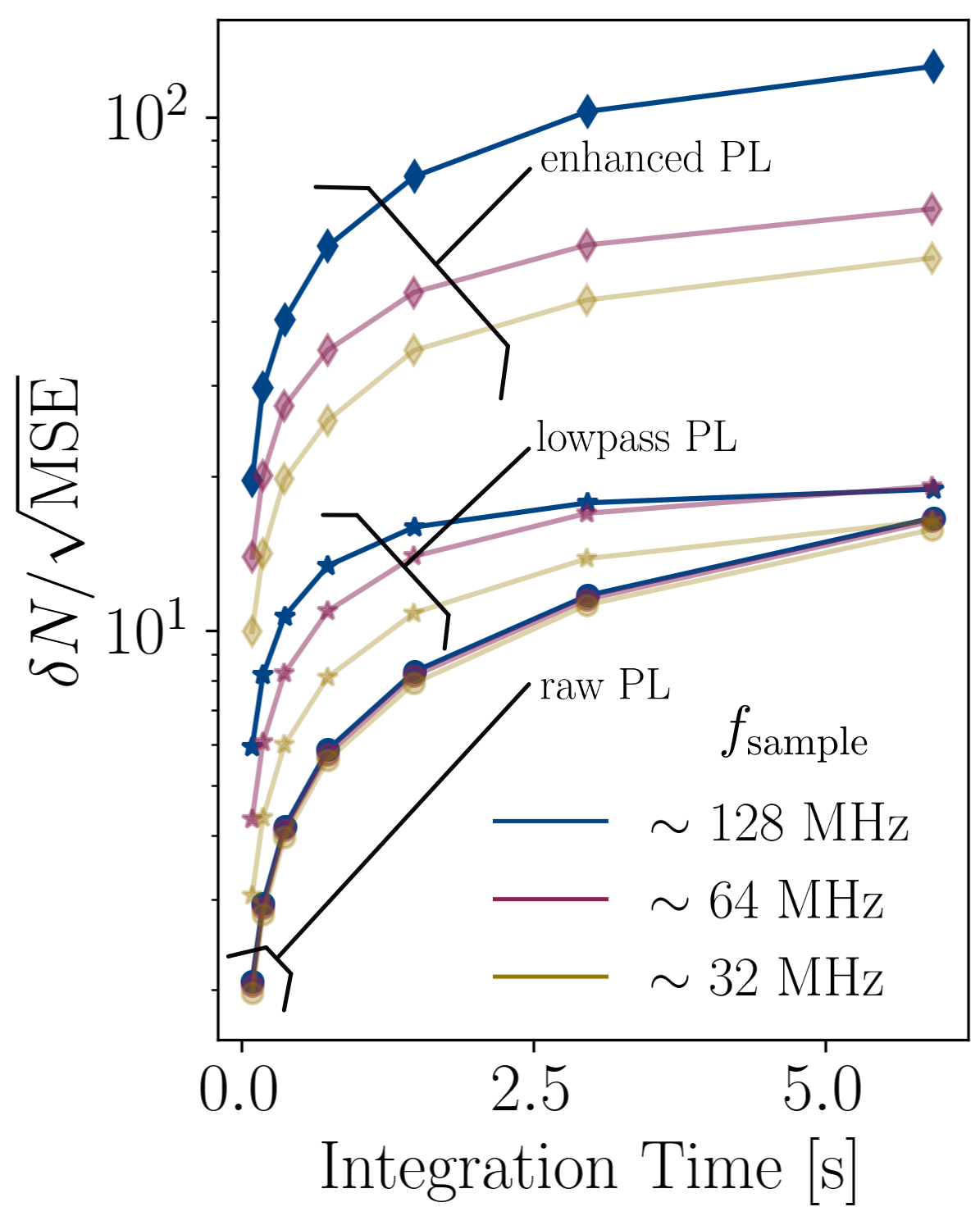}
  \caption{SNR with respect to integration time for PL acquired at various $f_{\text{sample}}$ values with fixed PL durations collected in the time interval $T_I = 3.5~\mu$s ($[t_i=0.2~\mu$s,  $t_f=3.7~\mu$s]) including 9 $n_{\text{SD}}$ points. Yellow, red, and blue lines represent the PL data with sampling frequencies approximately $32$~MHz, $64$~MHz, and $128$~MHz, respectively, as indicated by the insets. Lines with circle and star markers represent raw and lowpass filtered PL, respectively. Those with triangle markers indicate TMT-enhanced data obtained with optimum filter order.  }\label{SNR2}
\end{center}
\end{figure}

To put it into context, first, we would like to delineate the bias-variance trade-off by examining two extreme limits of the wavelet filter order, $\beta$.  The lower extreme $\beta \rightarrow -\infty$ corresponds to the raw PL limit (i.e., without wavelet denoising) where $N^{\beta}$ approaches to $N^{\text{raw}}$.  This regime governs low-bias/high-variance statistics for the random variable $N^{\beta}$. The upper extreme limit $\beta \rightarrow \infty$ corresponds to $N^{\beta} \rightarrow \mathcal{K} (t, \omega_{\text{temp}})$ which imposes high-bias/low-variance.  Thus, there is a bias-variance trade-off that should be optimized for the minimum MSE\@.  In Fig.~\ref{SNR} (a), we plot this trade-off as a function of filter order for various PLs generated with different $M$ values.  For low filter orders the squared bias is extremely small, while for high filter orders, MSE gets dominated by bias. The optimum order values, $\beta_{\text{opt}}$, (marked with dots in Fig.~\ref{SNR} (a)) minimize the MSE, and they get smaller for higher $M$, as expected from a consistent estimator.

In Fig.~\ref{SNR} (b) we display signal-to-noise ratio (SNR), defined as the signal amplitude due to the sensing field, $\delta N$, divided by the noise amplitude obtained as the square root of MSE with respect to the integration time. The scaling of SNR with respect to the integration time can be expressed as $c(MT_I)^\alpha$, where we refer to $c$ as the prefactor and $\alpha$ as the exponent.  The SNR lines constructed with different $M$ values for specific PL duration and fixed sampling frequency $f_{\text{sample}}$, possess different exponents, all of which are below that of SQL (i.e., $\alpha<0.5$).  Fortunately, despite the inferior exponent, filtered-SNR lines are exalted by very large prefactors resulting in significant performance gain. These scaling traits are the ramifications of the shrinkage of margin-exceeding points through hard thresholding. While pruning of data points to the closest margin value causes partial information loss, again the same pruning yields a large prefactor due to decreased variance and hence, reduced MSE overall.  Because of its $\alpha<0.5$ exponent, TMT-enhanced PL favors low-time-budget sensing applications. Nevertheless, for all practical purposes, this embraces very large integration times as seen by the crossing of their asymptotic behaviors (around $\sim 350$ seconds for the green line). 

\subsection{Comparison with frequency-based filtering}
To put into perspective the merits of wavelet filtering, we also employ a Butterworth lowpass filter to benchmark its SNR enhancements at different sampling rates. The Butterworth filter is renowned for maximally flat frequency response in the pass-band region and in our case, we used a sixth-order Butterworth filter with a cutoff frequency of $5$~MHz. It is possible to argue that a \textit{bandpass} filter might perform better when compared to a lowpass filter. However, in the rotating frame of reference, with $B_{\text{calib}}=100~\mu$T, the central frequency of the signal we would like to resolve is around $2.8$~MHz.  Considering the frequency resolution of approximately $0.28$~MHz implies that the central signal frequency is not well isolated from the DC wall, which hampers bandpass filter employment at a desirably high order. We make use of so-called forward-backward filtering to neutralize phase response and keep the slope detection points intact. In Fig.~\ref{SNR2}, we show SNR for varying $f_{\text{sample}}$ at fixed PL duration $T_I$.  The data points on each curve represent doubly ascending $M$ numbers, from 25K to 1.6M. As expected, for the raw PL changing the sampling frequency does not improve SNR, since it is the $M$ value that quantifies the sensitivity. For low $M$, the lowpass filter improves SNR effectively, yet, as $M$ increases lowpass filtered SNR values converge to unfiltered ones since an increasing number of uncorrelated measurements acts as a natural lowpass filter. The Butterworth filter performs better as increased $f_{\text{sample}}$ resolves high-frequency fluctuations in the low $M$ regime. For TMT-enhanced PLs an elevated number of data points provide more accurate $\omega_{\text{temp}}$ prediction for TMT-enhanced PLs which might become useful when sensing a large magnetic field in which PL needs to be sampled at a high rate to prevent phase wrapping \cite{degen17}.  Apart from obtaining significant improvement when compared to a lowpass filter, the TMT method also demonstrates its distinction in the high $M$ regime as well.

\subsection{Practical use of TMT}

\begin{figure}[tb]
  \begin{center}
  \includegraphics[width=1.0\columnwidth]{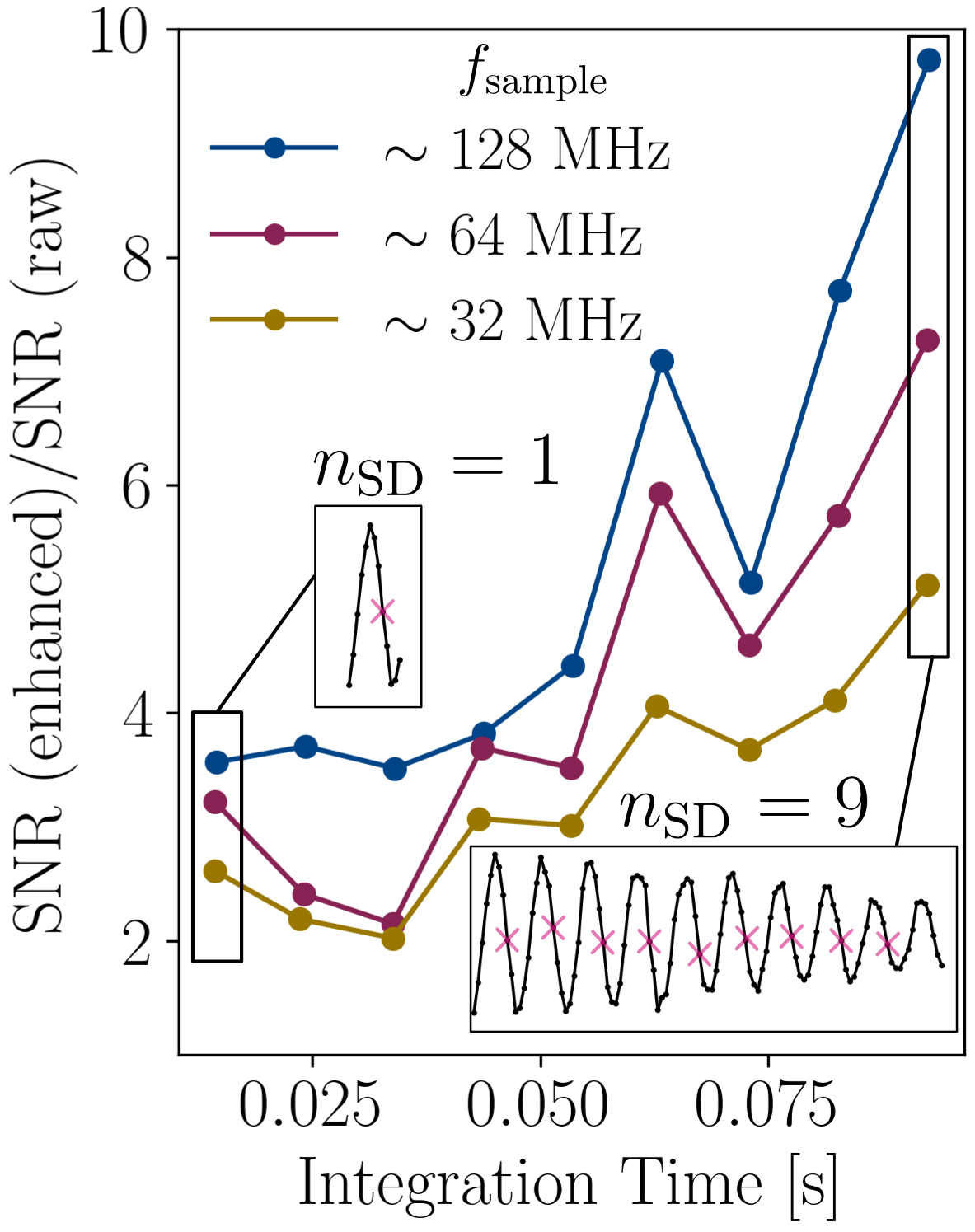}
  \caption{The performance gain from the TMT method for different PL durations
  at fixed $M =25$~K. The leftmost data point represents the gain in  PL
duration with only one detection point ($[t_i=0.2~\mu$s, $t_f=0.58~\mu$s]), and
the rightmost data point ($[t_i=0.2~\mu$s, $t_f=3.7~\mu$s]) is the longest PL
which includes nine $n_{\text{SD}}$ points as shown in the insets. In each data
point along the horizontal axis, PL is prolonged approximately by $0.38~\mu$s.
Yellow, red, and blue lines represent PLs collected with sampling frequencies
approximately $32$~MHz, $64$~MHz, and $128$~MHz, respectively. Insets show raw
PL data sampled at $32$~MHz. All of the TMT-enhanced data was obtained with the
filter order that minimizes the MSE of the calibration PL.}\label{SNR3}
\end{center}
\end{figure}

In general, TMT-filtered PL becomes an optimum estimate, when operated at $\beta_{\text{opt}}$ as it minimizes the MSE. However, in daylight applications, determining the optimal value for $\beta$ requires the knowledge of the true value of the number of photons at each fringe, $N^*(t_q)$, which is the parameter we would like to estimate in the first place. At this point, the calibration PL becomes handy so that the determination of filter order can be obtained from the calibration signal when there is no sensing field, $B_{\text{sense}}$.  Even though this value does not correspond to the optimum order for the particular sensing PL, as long as order-dependent parameters ($T_I, M, f_{\text{sample}}$) are the kept same, the optimum filter order for sensing protocol is observed to be close to the one in the calibration PL. To substantiate this claim, in Fig.~\ref{SNR3}, we operate at the filter order that minimizes the MSE of the calibration PL (unlike those in Figs.~\ref{SNR2} and \ref{SNR3}). To directly assess its performance improvement, SNR of TMT-enhanced-PL is normalized to that of raw-PL, \emph{i.e.}, $\sqrt{\text{MSE}[N^{\text{raw}}]/ \text{MSE} [N^{\beta}}]$, where $\text{MSE}[N^{\text{raw}}]$ is the MSE of the PL before enhancement.  Here, we also compare three different sampling regimes. Increasing the duration of PL at a fixed sampling rate and $M = 25$K, in general, refines the performance as the template can be estimated more accurately so that the filter order employed, $\beta$, can be larger. The non-monotonic performance improvement predominantly originates from the fact that the normalized MSE is calculated empirically from a rather small sample size of $n_{\text{exp}}=2000$, rendering it prone to statistical fluctuations.

\subsection{General Traits}
Overall, with long PL acquisitions and reasonable $M$ values, it is possible to reach a significant PL enhancement up to an order of magnitude for each slope detection point on average.  As $M\rightarrow \infty$, the template frequency, $\omega_{\text{temp}}$ approaches the true value $\omega^*$, and hence the filtered PL converges to the true PL providing a sufficient condition for being a \textit{consistent} estimator.

In principle, there is no fundamental barrier to performance improvement via TMT, other than the $T_2^*$ decay factor. The reason is that the template frequency, $w_{\text{temp}}$ which is determined from the maximizing value of $R(\omega)$, prioritizes the early fringes as PL amplitude diminishes due to decoherence such that MSE in $q$th fringe is less than MSE of the fringes $q+1, \dots, n_{\text{SD}}$ for the same PL and filter order, $\beta$. Furthermore, prolonging the PL duration after some $t_f$ is not going to be useful due to diminishing contrast between $n_0$ and $n_1$.  In this respect, TMT-boosted performance would be even higher for the isotopically purified NV-center with $T_2^*$ reaching $90~\mu$s \cite{ishikawa12}, or by combining with other measures like dynamical decoupling pulse sequences~\cite{barry20}. 

\section{Conclusion}\label{conclusion}
In summary, we present a wavelet-based PL denoising technique that can increase the SNR by around an order of magnitude with respect to raw data.  This improvement's functional dependence on the integration time is dominated by the large prefactor as opposed to the scaling exponent. Being applied as a post-processing over the PL data, TMT has practical simplicity and versatility.  Our proof of principle computational assessment relied on a single NV center DC magnetometry. However, we also think that the method holds great promise for ensemble sensing and for various time-dependent fields.

\section*{Acknowledgement}
This work was supported by Wavelet-Enhanced Quantum Sensing with Solid-State
Nuclear Spins (AFOSR\,FA9550-22-1-0444).
The numerical calculations reported in this paper were partially performed at
T\"UB\.ITAK ULAKB\.IM, High Performance and Grid Computing Center (TRUBA
resources). We thank Mustafa G\"{u}ndo\u{g}an for fruitful discussions.
\section*{Appendix}
In this appendix, we first intend to give some preliminary information about wavelets to bridge with the material in Sec.~\ref{TMT}. Then, we present a mathematical description of how TMT shrinkage should be executed within the wavelet basis. Lastly, we speculate on why scaling of TMT enhancement is in the sub-SQL regime. 
\subsection{Continuous wavelet transform}
The continuous wavelet transform of a square integrable signal, $x(t)$, can be written as,
\begin{equation}
  D(a, b) = \dfrac{1}{\sqrt{a}}\int^{\infty}_{-\infty} x(t)\, \psi^* \left(\dfrac{t-b}{a}\right)dt,
  \end{equation}
where $a$ and $b$  denote the scale (dilation) and translation parameters of the wavelet function, and $*$ represents its complex conjugation.  From a mathematical point of view, we convolve $x(t)$ with some $\psi(t)$ which is localized both in the time and frequency domain, instead of sinusoidal functions that have frequency localization only as in Fourier transform.  This enables time-frequency localization of the $x(t)$ simultaneously.  It is important to note that, the complete representation of $x(t)$ requires the knowledge of $D(a,b)$ for each infinitely small step of $a$ and $b$ which constitutes a highly redundant representation of the signal. 

\subsection{Continuous time discrete wavelet transform}
It is possible to provide a more compact representation of the original signal by discretizing the scale and translation parameters of the wavelet function as, 
\begin{equation}
\psi_{j,k}=\psi\left(\dfrac{t-kb_0a_0^j}{a_0^j}\right)/\sqrt{a_0^j}\, , 
\end{equation}
where $j,k \in Z$. If we consider a dyadically-grided scale and translation parameters ($a_0=2$ and $b_0=1$, as an example) provided that, $\psi_{j,k}$ constitutes a \textit{sufficiently} tight frame \cite{daubechies92}, the function $x(t)$ can be expressed in linear combinations of wavelets at different scales ($j$) and locations ($k$),
\begin{equation}
x(t) = \sum_j\sum_k d_{j,k} \psi_{j,k}.
  \end{equation}
where $d_{j,k}= \langle x(t), \psi_{j,k}(t)  \rangle = \int_{-\infty}^{\infty} x(t) \psi_{j,k}(t)dt $.  For $x(t)$ with non-vanishing mean the summation over $j$ requires infinitely many terms, since, $\int \psi_{j,k}(t)dt=0$. This can be visualized in the frequency domain as indicated in Fig. 1~(b), increasing $j$ by one unit, just halves the bandwidth of the wavelet which has no zero-frequency component for any finite value of $j$. To overcome this difficulty,  another set of functions, called scaling functions $\phi_{j,k}$, can be introduced with non-zero mean. Scaling functions are orthogonal to wavelets at the same scale, yet they form a basis for the wavelets at the lower scale, for instance, the so-called mother wavelet can be expressed as~\cite{walker08},
\begin{equation}\label{m2f}
  \psi_{0,0}(t)=\sum_k h_1(k)\sqrt{2}\phi_{0,0}(2t-k),
\end{equation}
where $h_1(k)$ is some coefficient depending on the wavelet type, and $\phi=\phi_{0,0}$ is termed as the father wavelet as an analog to the mother wavelet. Father wavelet can be utilized to create scaling functions at different scales and locations as $\phi_{j,k} = 2^{-j/2}\phi(2^{-j}t-k)$. Then, function $x(t)$ can be expressed more conveniently at given scale $J$ as,
\begin{equation}
  x(t) = \sum_{k=-\infty}^{\infty}c_{J, k}\phi_{J,k}(t) + \sum_{j=0}^{J}\sum_{k=-\infty}^{\infty} d_{j,k}\psi_{j,k}(t),
\end{equation}
$c_{J, k}=\langle x(t), \phi_{J,k}(t)\rangle$  and $d_{j, k}=\langle x(t), \psi_{j,k}(t)  \rangle$ are called as approximation and detail coefficients, respectively. 
\subsection{Discrete time fast wavelet transform and undecimated wavelet transform}

For discrete-time signals,  the DWT can be calculated by half-band-lowpass and half-band-highpass filters. Since the lower or higher half of the frequencies are removed for the signal of interest the output discrete-time signals can be decimated by two without loss of information according to Nyquist sampling theorem, so that,
\begin{align} 
  c_{j,k} &= \sum_m h_0(m-2k)c_{j-1, m}, \label{h0}\\
  d_{j,k} &= \sum_m h_1(m-2k)c_{j-1, m}, \label{h1}
\end{align}
where, $c_{j,k}$ and $d_{j,k}$ are approximation and detail coefficients at the $j$th detail level, respectively. $h_0$ ($h_1$) is  half-band-lowpass (half-band-highpass) filters and the index $2k$ implements the decimation provided that the filters $h_0$ and $h_1$ satisfy the properties of  quadrature mirror filters and they are the coefficients connecting the wavelet and scaling functions at different scale as Eq.~(\ref{m2f}) and,
\begin{equation}
  \phi_{0,0}(t)=\sum_k h_0(k)\sqrt{2}\phi_{0,0}(2t-k).
\end{equation}

With the help of Eqs.~($\ref{h0}$) and ($\ref{h1}$) it is possible to find a \textit{compact} representation of a signal of interest in the wavelet domain.  However, decimation breaks the time-invariance which makes the TMT-like filtering method very challenging. Therefore, by moving into a more \textit{redundant} representation with (\textit{i.e. undecimation}) one can restore the time invariance. The DWT with undecimation is sometimes called a Stationary Wavelet Transform (SWT) or Redundant Wavelet Transform along with the Undecimated Wavelet Transform (UWT) as we used throughout this paper.

\subsection{TMT-shrinkage}\label{shrink}

The UWT of upper and lower margins in the wavelet domain can be calculated up to a maximum detail level $J$ as,
\begin{align}
  \text{UWT}(\mathcal{U}) &= [d_0^{\mathcal{U}}, d_1^{\mathcal{U}}, \dots, d_J^{\mathcal{U}}, c_J^{\mathcal{U}}], \\
  \text{UWT}(\mathcal{L}) &= [d_0^{\mathcal{L}}, d_1^{\mathcal{L}}, \dots, d_J^{\mathcal{L}}, c_J^{\mathcal{L}}],
\end{align}
from which  time-dependent margins can be constructed for the $j$th detail level $l_{j,k}=\min(d_{j,k}^{\mathcal{U}}, d_{j, k}^{\mathcal{L}})$ and $u_{j, k}=\max(d_{j, k}^{\mathcal{U}}, d_{j, k}^{\mathcal{L}})$ as marked in Fig.~$\ref{algo}$ (d). We choose margins for the allowed signal range with respect to $l_{j, k}, u_{j, k}$, such that,
\begin{equation} \label{hard}
  R_{j, k}^{\text{hard}} =
    \begin{cases*}
      d_{j, k}^{\text{raw}},  & $l_{j, k} \le d_{j, k}^{\text{raw}} \le u_{j, k}$ \\
      u_{j, k}^{\mathcal{K}}, & $d_{j, k}^{\text{raw}}> u_{j, k}$ \\
      l_{j, k}^{\mathcal{K}}, & $d_{j, k}^{\text{raw}}< l_{j, k}$
    \end{cases*},
  \end{equation}
where $R_j^{\text{hard}}$ and $d_j^{\text{raw}}$ denote the filtered  and raw PL at the $j$th decomposition level, respectively, for hard wavelet shrinkage (Fig.~\ref{algo}~(d)).  We note that there are various shrinkage types available~\cite{nason08}, yet in this paper, we only use hard wavelet shrinkage for simplicity. By obtaining filtered PL for each detail level, one can recover the enhanced PL in the time domain through inverse UWT, that is,
\begin{equation}
N^{\beta} =  \text{IUWT}[R_0, R_1, \dots, R_J, c_J^{\text{raw}}], \\
\end{equation}
where $N^{\beta} $ represents the enhanced sensing PL signal with the TMT method using hard shrinkage and IUWT denotes inverse undecimated wavelet transform.
\subsection{Some remarks on sub-SQL scaling of TMT}
Considering $M$ number of repeated measurements at fixed PL collection duration, $T_I$, focusing on the statistics of a single slope detection point of the PL to keep the calculation simple the estimator $\hat{N}$ can be constructed from sample mean,
\begin{equation}
  \hat{N} = \dfrac{1}{M}\sum_{i=1}^{M} N_i,
\end{equation}
where $N_i$ represents the number of photons in $i$th statistical observation at the slope detection point.  Then the total error in the estimate can calculated as the square root of MSE.  That is,
\begin{align*}
  \text{MSE}[\hat{N}] &= \text{Var}\left[\hat{N}\right] + \text{Bias}\left[\hat{N}\right]^2, \\
                      &= \text{Var}\left[\dfrac{1}{M}\sum_{i=1}^M N_i\right] +  \text{Bias}\left[\dfrac{1}{M}\sum_{i=1}^M N_i\right]^2, \\
                      &= \dfrac{1}{M^2} \text{Var}\left[\sum_{i=1}^M N_i\right] + \text{Bias}\left[\dfrac{1}{M}\sum_{i=1}^M N_i \right]^2, \\
\end{align*}
For the raw PL, the estimator is unbiased so that the square root of MSE reduces to,
\begin{equation*}
  \sqrt{\text{MSE}} = \dfrac{1}{M}\sqrt{\sum_{i=1}^M \text{Var}[N_i]} = \dfrac{1}{M} \sqrt{M\sigma^2_{\text{raw}}} = \dfrac{\sigma_{\text{raw}}}{\sqrt{M}}
\end{equation*}
which is known as the standard error where we assume that $\text{Var}[N_i]=\sigma_{\text{raw}}^2$ for all $i=1,\dots, M$ implying a SQL scaling ($\sqrt{M}$).  For successfully enhanced PL, the standard deviation is strictly less than that of the raw PL and is unlikely to be the reason behind sub-SQL scaling (\textit{i.e.}  $\sigma_{\text{TMT}}<\sigma_{\text{raw}}$, where $\sigma_{\text{TMT}}$ is the standard deviation for enhanced data). On the other hand, for TMT-enhanced PL, we have non-vanishing bias term $\epsilon = \text{Bias}\left[\sum_{i=1}^M N_i / M\right]$ which can be expressed as a function of $M$ with some exponent $\gamma$, such that, $\epsilon(M) = \epsilon_0M^{\gamma}$. Then,
\begin{align*}
  \sqrt{\text{MSE}} &= \dfrac{1}{M} (M\sigma_{\text{TMT}}^2 + \epsilon_0^2 M^{2\gamma} )^{1/2}, \\
                    &= M^{\gamma-1} (M^{1-2\gamma}\sigma_{\text{TMT}}^2 + \epsilon_0^2 )^{1/2}.
\end{align*}
To be responsible for such a sub-SQL scaling (see in Fig.~\ref{SNR} (b)), the exponent of the bias term must be $0.5<\gamma<1$. Considering that the $M$ values are quite large, the above expression reduces to $M^{\gamma-1}\epsilon_0$ which in turn corresponds to sub-SQL scaling as $(1-\gamma)<0.5$.

%

\end{document}